# Crossover from Boltzmann to Wigner thermal transport in thermoelectric skutterudites

Enrico Di Lucente[1,*] Michele Simoncelli[2] and Nicola Marzari[1,3]

[1]*Theory and Simulation of Materials (THEOS) and National Centre for Computational Design and Discovery of Novel Materials (MARVEL), École Polytechnique Fédérale de Lausanne, Lausanne 1015, Switzerland*
[2]*TCM Group, Cavendish Laboratory, University of Cambridge, 19 JJ Thomson Avenue, Cambridge, CB3 0HE, United Kingdom*
[3]*Laboratory for Materials Simulations, Paul Scherrer Institut, 5232 Villigen PSI, Switzerland*

Skutterudites are crystals with a cagelike structure that can be augmented with filler atoms ("rattlers"), usually leading to a reduction in thermal conductivity that can be exploited for thermoelectric applications. Here, we leverage the recently introduced Wigner formulation of thermal transport to elucidate the microscopic physics underlying heat conduction in skutterudites, showing that filler atoms can drive a crossover from the Boltzmann to the Wigner regimes of thermal transport, i.e., from particlelike conduction to wavelike tunneling. At temperatures where the thermoelectric efficiency of skutterudites is largest, wavelike tunneling can become comparable to particlelike propagation. We define a Boltzmann deviation descriptor able to differentiate the two regimes and relate the competition between the two mechanisms to the materials' chemistry, providing a design strategy to select rattlers and identify optimal compositions for thermoelectric applications.

## I. INTRODUCTION

Heat is a waste product of many and diverse energy intensive technologies, from vehicle exhausts in transportation to nuclear and natural gas power plants production, to large-scale manufacturing. Ongoing research is focused on finding strategies to convert waste heat into electricity, and thermoelectric materials are among the most promising candidates [1] for this task, and for augmenting sustainable energy supplies in the near future. The thermoelectric figure of merit reaches a record value of 3.1 at 783 K in polycrystalline SnSe [2], even greater than the value obtained for single crystals [3] due to enhanced scattering. While many efforts have focused on designing materials with enhanced thermoelectric performance [4,5], understanding how to maximize energy-conversion efficiency by decreasing thermal conductivity has been hindered by the lack of a microscopic theory capturing the mechanisms of heat conduction in poor thermal conductors. Fittingly, the recently developed Wigner formulation of thermal transport [6,7] allows to describe heat conduction in anharmonic crystals and in solids with ultralow or glasslike conductivity; this is exactly the case of thermoelectrics. The Wigner formulation offers a comprehensive approach to describe heat transport across different regimes, covering on the same footing harmonic "Boltzmann," anharmonic "Wigner" and amorphous solids and glasses regimes. A crystal exhibiting Boltzmann thermal transport is characterized by phonon-interband spacings much larger than the phonon linewidths; in these materials particlelike heat conduction dominates [6–8] and the Peierls-Boltzmann transport equation [9,10] describes accurately thermal conductivity [11,12]. In amorphous solids and glasses wavelike tunneling dominates and the Wigner formulation recovers the Allen-Feldman formulation [13] accounting also for anharmonicity [14]. Last, crystals exhibiting Wigner thermal transport can be considered as the intermediate regime between the first two and are characterized by interband spacings which are comparable to phonon linewidths. The Peierls-Boltzmann equation fails to describe the wavelike contributions [15–18] in materials with ultralow thermal conductivity, that are captured by the Wigner transport equation [6,7].

Among others, skutterudites have been extensively studied for their possible applications in thermoelectrics [19–21] showing both high electrical and low thermal conductivities. The cagelike structure of skutterudites is a critical feature for thermoelectricity [21,22]: the voids present in the structure can be occupied by loosely bound atoms ("fillers" or "rattlers") that can reduce thermal conductivity and enhance the thermoelectric figure of merit [23]. However, the interpretation of filler vibrations remains not fully understood [16,24,25]. The explanation of the rattling motion based on filler vibrations concentrated in specific energy ranges [24] was contradicted by a theoretical study suggesting the presence of a strong hybridization between the vibrations of the filler atoms and specific phonon bands of cage atoms [25]. This concept was then generalized to that of a "coherent interaction between fillers and host matrices" [16], which is more intricate than the simplistic "bare rattling." Here, we emphasize that also the latter explanation did not investigate the role of the filler in terms of phonon wavelike heat conduction and so we aim to clarify the fundamental mechanisms behind the reduction of thermal conductivity—and thus the related thermoelectric performance—applying the Wigner transport equation.

---

*enrico.dilucente@epfl.ch

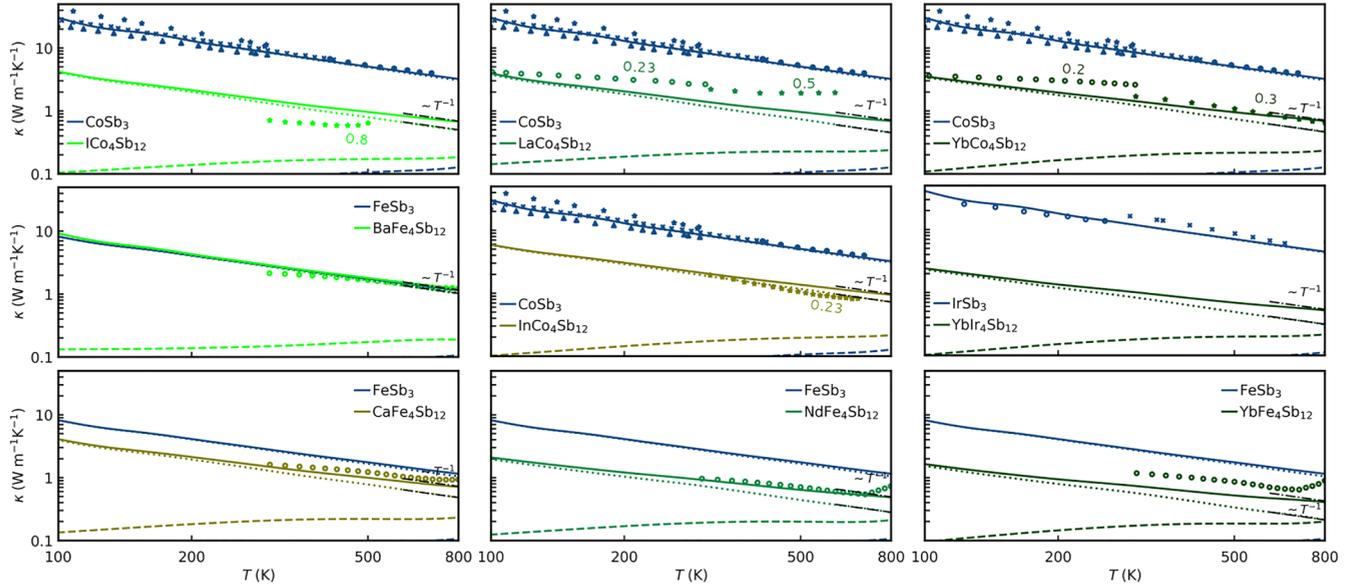

FIG. 1. Calculated $\kappa_{\text{tot}}$ (solid), $\kappa_P$ (dotted), and $\kappa_C$ (dashed) for unfilled FeSb$_3$, CoSb$_3$, and IrSb$_3$ and for their related filled compounds RFe$_4$Sb$_{12}$ (R = Ba, Ca, Nd, Yb), RCo$_4$Sb$_{12}$ (R = I, In, La, Yb), and RIr$_4$Sb$_{12}$ (R = Yb), between 100 and 800 K. Experimental $\kappa_{\text{tot}}$ are taken from Ref. [34] for RFe$_4$Sb$_{12}$ (R = Ba, Ca, Nd, Yb), Refs. [35,36] for CoSb$_3$, Refs. [37,38] for ICo$_4$Sb$_{12}$, Ref. [39] for InCo$_4$Sb$_{12}$, Refs. [40,41] for LaCo$_4$Sb$_{12}$, Refs. [42,43] for YbCo$_4$Sb$_{12}$, and Refs. [44,45] for IrSb$_3$. Symbols represents different experimental measurements, also color-coded according to the material. Experimental data referring to partial concentration of the filler are specified adjacent to the symbols. The universal $T^{-1}$ trend of $\kappa_P$ [16,46] is also shown; $\kappa_{\text{tot}}$ displays a milder decay than $\kappa_P$.

## II. METHOD

The Wigner formulation of thermal transport [6,7] yields the following thermal conductivity expression:

$$\kappa_{\text{tot}}^{\alpha\beta} = \kappa_{\text{P,SMA}}^{\alpha\beta} + \frac{1}{(2\pi)^3} \int_{\text{BZ}} \sum_{s \neq s'} \frac{\omega(\boldsymbol{q})_s + \omega(\boldsymbol{q})_{s'}}{4}$$
$$\times \left[ \frac{C(\boldsymbol{q})_s}{\omega(\boldsymbol{q})_s} + \frac{C(\boldsymbol{q})_{s'}}{\omega(\boldsymbol{q})_{s'}} \right] v^\alpha(\boldsymbol{q})_{s,s'} v^\beta(\boldsymbol{q})_{s',s}$$
$$\times \frac{\frac{1}{2}[\Gamma(\boldsymbol{q})_s + \Gamma(\boldsymbol{q})_{s'}]}{[\omega(\boldsymbol{q})_{s'} - \omega(\boldsymbol{q})_s]^2 + \frac{1}{4}[\Gamma(\boldsymbol{q})_s + \Gamma(\boldsymbol{q})_{s'}]^2} \, d^3q, \quad (1)$$

where $\omega(\boldsymbol{q})_s$ is the angular frequency of a phonon having wave vector $\boldsymbol{q}$ and mode $s$, $C(\boldsymbol{q})_s = \frac{\hbar^2 \omega^2(\boldsymbol{q})_s}{k_B T^2} \tilde{N}(\boldsymbol{q})_s [\tilde{N}(\boldsymbol{q})_s + 1]$ is the specific heat of that phonon mode, $\tilde{N}(\boldsymbol{q})_s$ is the equilibrium Bose-Einstein distribution at temperature $T$, $v^\alpha(\boldsymbol{q})_{s,s'}$ and $v^\beta(\boldsymbol{q})_{s,s'}$ are the cartesian components of the velocity operator, which generalizes the concept of group velocity, and $\Gamma(\boldsymbol{q})_s = \frac{1}{\tau(\boldsymbol{q})_s}$ is the phonon linewidth of a phonon with lifetime $\tau(\boldsymbol{q})_s$. The symbol BZ in Eq. (1) represents an integral over the Brillouin zone. $\kappa_{\text{P,SMA}}^{\alpha\beta}$ in Eq. (1) is the Peierls-Boltzmann particlelike conductivity, which is driven by phonon-phonon scattering in the single-mode relaxation time approximation (SMA). The additional Wigner term in Eq. (1) is a positive-definite tensor ($\kappa_C^{\alpha\beta}$) emerging from the phase "coherences" between pairs of phonon eigenstates; i.e., from the wave-tunneling between two nondegenerate bands

($s \neq s'$) [26,27]. To explore the relative strength of the particlelike and wavelike heat-conduction mechanisms, we study three different families of skutterudites, where we compute from first-principles (see Appendices for details) all the quantities needed to evaluate Eq. (1) for unfilled FeSb$_3$, CoSb$_3$, and IrSb$_3$, and for the filled compounds RFe$_4$Sb$_{12}$ (R = Ba, Ca, Nd, Yb), RCo$_4$Sb$_{12}$ (R = I, In, La, Yb) and RIr$_4$Sb$_{12}$ (R = Yb). We note, in passing, that skutterudites contain elements for which DFT+U [29–32] might improve upon self-interaction errors for localized electrons [33]; these aspects are discussed in Appendix D.

## III. DISCUSSION

The resulting thermal conductivities, together with their good agreement with the available experimental data, are shown in Fig. 1 in the temperature range from 100 to 800 K. By focusing on the high-performance high-temperature regime ($T \geqslant 600$K) and comparing $\kappa_P$ and $\kappa_C$, it is clearly seen how unfilled skutterudites are in the Boltzmann regime, displaying dominant particlelike conduction. Moreover, in the high-temperature region, $\kappa \simeq \kappa_P \propto T^{-1}$, as predicted by the Peierls-Boltzmann equation [46,47]. However, filled skutterudites display Wigner regime of thermal transport, with a milder decay of $\kappa_{\text{tot}}$ as also observed for many highly anharmonic crystals [16,17,48]. Among the materials studied, we highlight how Yb-filled materials show the most evident Wigner regime, i.e., $\kappa_P \sim \kappa_C$, while for BaFe$_4$Sb$_{12}$ the behavior is more close to the Boltzmann regime, i.e., $\kappa_P > \kappa_C$. It is worth noting that some experimental measurements show an increase in conductivity at very high temperatures ($\geqslant 700$ K); this could be understood as driven by radiative and electronic heat transfer [49,50], unrelated to the increase in coherences' conductivity. At high temperatures, radiative effects on thermal transports are expected to be important because radiative thermal conductivity should increase as $T^3$ [51], while in semiconductors the electronic contribution can be important when, as temperature rises, electrons are excited to the conduction band [50].

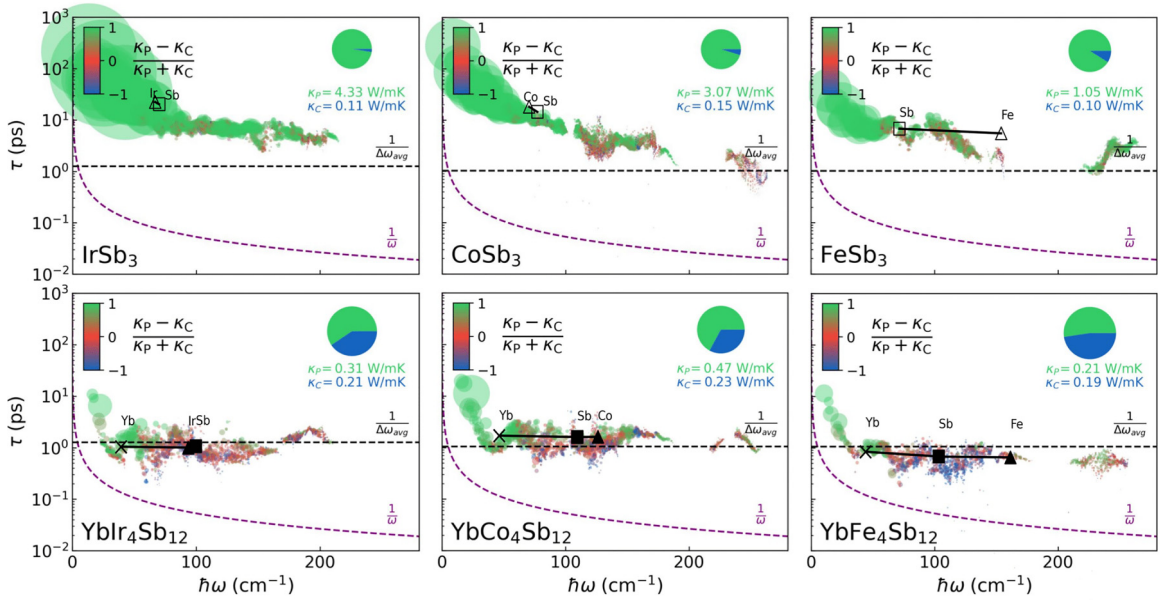

FIG. 2. Distribution of phonon lifetimes $\tau(\boldsymbol{q})_s = \Gamma(\boldsymbol{q})_s^{-1}$ as a function of energy $\hbar\omega(\boldsymbol{q})_s$ for unfilled (upper panels) and Yb-filled (lower panels) skutterudites at 800 K. The area of each scatter point is proportional to the contribution to the total lattice thermal conductivity and colored according to the origin of the contribution: $c = [\bar{\kappa}_P(\boldsymbol{q})_s - \bar{\kappa}_C(\boldsymbol{q})_s]/[\bar{\kappa}_P(\boldsymbol{q})_s + \bar{\kappa}_C(\boldsymbol{q})_s]$, where particlelike is green (c = 1), wavelike is blue (c = −1) and intermediate mechanisms have intermediate colors, with red corresponding to 50% of particlelike and 50% of wavelike contributions [7]. The Wigner limit in time (dashed-black line) corresponds to a phonon lifetime equal to the inverse of the average interband spacing ($\tau^\omega = \Delta\omega_{avg}^{-1}$). The dashed-purple hyperbola shows the Ioffe-Regel limit in time [7,28] ($\tau^{IR} = \omega^{-1}$), below which phonons are no longer well-defined quasi-particles. The pie charts have an area proportional to the total lattice thermal conductivity, and the slices resolve the particlelike conductivity (green) and the wavelike conductivity (blue). The black symbols connected by black lines are the points whose coordinates are the average energies and lifetimes projected on atoms [see Eqs. (A3) and (A4)]. Open (closed) symbols refer to unfilled (filled) skutterudites. The projections on the filler, transition-metal and antimony atoms are given by crosses, triangles and squares, respectively. The phonon lifetimes distribution for the remaining filled skutterudites are given in Appendix F 2.

To elucidate the physics underlying the low thermal conductivity of efficient thermoelectric skutterudites, we make a comparison between phonon lifetimes and average phonon interband spacing $\Delta\omega_{avg} = \frac{\omega_{max}}{3N_{at}}$ ($\omega_{max}$ being the maximum phonon frequency and $N_{at}$ the number of atoms in the primitive cell) [7], to describe how much each phonon contributes to the wavelike vs. the particlelike conduction mechanisms. As given in Ref. [7] (see Appendix E of Ref. [7] for a detailed derivation), it is possible to resolve how much each phonon $(\boldsymbol{q})_s$ contributes to the particlelike and wavelike conductivities. The ratio between the two is approximately equivalent to the ratio between the phonon linewidth and the average phonon interband spacing,

$$\frac{\bar{\kappa}_C(\boldsymbol{q})_s}{\bar{\kappa}_P(\boldsymbol{q})_s} \simeq \frac{\Gamma(\boldsymbol{q})_s}{\Delta\omega_{avg}} = \frac{1}{\tau(\boldsymbol{q})_s \Delta\omega_{avg}}, \quad (2)$$

where $\bar{\kappa}_P(\boldsymbol{q})_s$ and $\bar{\kappa}_C(\boldsymbol{q})_s$ are the mode-dependent average trace of the thermal conductivity tensor for particlelike and wavelike heat conduction, respectively. From this one can define the Wigner limit in time $\tau^\omega = \frac{1}{\Delta\omega_{avg}}$, that determines the nonsharp crossover from a regime of dominant particlelike conduction to one of dominant wavelike conduction.

In Fig. 2 we show the distribution $\tau(\boldsymbol{q})_s$ of the phonon lifetimes as a function of the energy $\hbar\omega(\boldsymbol{q})_s$ for the unfilled (upper panels) and Yb-filled (lower panels) skutterudites at 800 K. Phonons above the Wigner limit in time [i.e., with $\tau(\boldsymbol{q})_s > \tau^\omega$] contribute mainly to particlelike conductivity, while phonons below this limit [i.e., with $\tau(\boldsymbol{q})_s < \tau^\omega$] contribute mainly to the wavelike conductivity. We note that in all the compounds studied (see Appendix F 2) the phonon lifetimes sit well above the Ioffe-Regel limit in time $\tau^{IR} = \omega^{-1}$ [7] (dashed-purple), confirming that phonons in these materials are well defined quasi particles and that the Wigner formulation is valid [7]. If this were not the case, then full spectral function approaches [52,53] would be required. The clouds of phonon lifetimes in Fig. 2 remain distinctly above the Wigner limit for the unfilled compounds, while they are centered around the Wigner limit for the Yb-filled compounds. This allows to identify the crossover from the Boltzmann regime of unfilled skutterudites to the Wigner regime of Yb-filled ones, where phonon coherences become significant. Moreover, Fig. 2 also shows the atom-resolved energies and lifetimes [see Eqs. (A3) and (A4)], indicating how, in general, different atoms contribute to different regions of the distribution. We see that Yb fillers drive the characteristic lifetimes toward the Wigner limit in time, while the characteristic energies of the Ir/Co/Fe atoms, and of Sb are only slightly shifted higher. Given a certain phonon lifetime, the frequency associated to the filler is significantly lower than that of the other atom types in the structure, suggesting that indeed Yb behaves like a rattler [54]. These findings extend those discussed in Ref. [16], where it was suggested that fillers interact with the host matrix coherently, albeit without resolving the single-atom contributions in the energy-lifetime distribution. In fact, the present analysis shows how filling with Yb lowers all the lifetimes, not only those of the filler itself. Moreover, from Eq. (2) we see that when the wavelike contribution to the thermal conductivity is much bigger than the particlelike one, $\kappa_C \gg \kappa_P$, the linewidths are comparable to the phonon interband spacing. We also highlight that in the long-wavelength limit $\mathbf{q} \to 0$ the linewidths of the optical modes $\Gamma(\omega)_s$ with $s > 3$ remain finite (and, as shown in Fig. 2, smaller than $\omega$, ensuring that the quasiparticle picture is well defined) and the optical phonon bands flatten in proximity of $\mathbf{q} = 0$; these conditions imply that when $\mathbf{q} \to 0$ optical phonons contribute exclusively to the wavelike conductivity. In contrast, for acoustic phonons the linewidths $\Gamma(\omega)_s$ with $s = 1, 2, 3$ tend to zero as $\omega \to 0$—as shown in Fig. 2, the linewidths tend to zero faster than $\omega$, ensuring that the quasiparticle (nonoverdamped) condition $\omega > \Gamma(\omega)_s$ is respected—so the overlap between different phonon bands is negligible and the wavelike contribution of acoustic phonons to the conductivity is negligible. This implies that, while approaching $\mathbf{q} \to 0$, the acoustic modes contribute particlelike to the conductivity, and their zero specific heat at $\mathbf{q} = 0$ implies that they contribute zero at the limiting point $\mathbf{q} = 0$. The particlelike behavior of the acoustic modes when approaching $\mathbf{q} \to 0$ can be intuitively understood. In fact, we recall that the lifetime of acoustic phonons is mainly determined by anharmonic phonon-phonon collisions, which are totally inelastic scattering events as they do not conserve the number of phonons. Thus, the scattering time $1/\Gamma(\omega)_s$ of nonoverdamped acoustic phonons coincides with their decoherence time—a condition that, in addition to the lack of significant overlap with different phonon bands, is indicative of particlelike behavior [55].

Then, we want to describe how each chemical species that composes the skutterudite structure influences the relative strengths of $\kappa_P$ and $\kappa_C$. Since the Boltzmann or Wigner regime is regulated by the competition between the phonon lifetimes and the Wigner limit in time [7] [see Eq. (2)], we introduce a Boltzmann deviation descriptor ($B$) defined as the inverse of the product between the skutterudite characteristic lifetime—i.e., the Matthiessen' sum of the average lifetimes resolved on atom types (see Appendix A for details)—and the average phonon interband spacing [7]:

$$B = \frac{1}{\bar{\tau}\,\Delta\omega_{\text{avg}}} = \begin{cases} \frac{1}{\tau^U \Delta\omega_{\text{avg}}}, & \text{with } \tau^U = \frac{\tau_M \tau_{Sb}}{\tau_M + \tau_{Sb}}, \\ \frac{1}{\tau^F \Delta\omega_{\text{avg}}}, & \text{with } \tau^F = \frac{\tau_R \tau_M \tau_{Sb}}{\tau_R \tau_M + \tau_R \tau_{Sb} + \tau_M \tau_{Sb}}, \end{cases} \quad (3)$$

where $\bar{\tau}$ is the average lifetime (A4), F and U superscripts symbolize filled and unfilled skutterudites, R the rattler, and M the Ir/Co/Fe metal. In the strategy for obtaining an average lifetime we first perform a convolution of the lifetimes of each phonon mode into a frequency-dependent representation [see Eq. (A1)]. Then we exploit the phonon partial density of states (PDOS) for evaluating the contribution coming from each atomic species in the material [see Eq. (A4)]. In this way, being able to take into account the lifetime of vibrations of the different atom types present in the structure (namely, filler and cage atoms), $B$ is able to quantify the deviation from the Boltzmann regime of thermal transport induced by fillers.

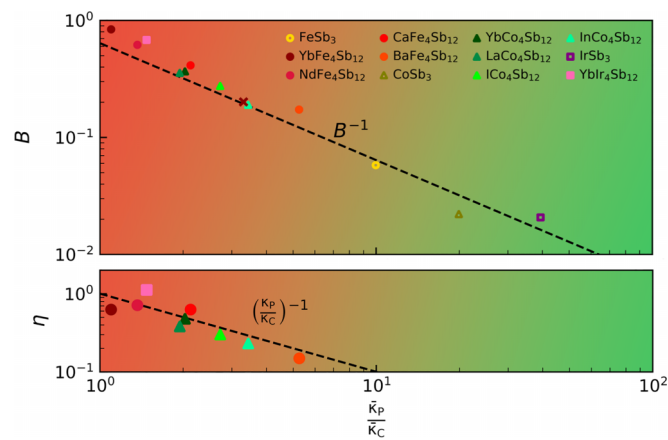

FIG. 3. (Upper panel) Relation between the relative strength of particlelike and wavelike conduction, $\frac{\kappa_P}{\kappa_C}$, at 800 K and $B$ as given in Eq. (3); unfilled (filled) symbols represent unfilled (filled) skutterudites, respectively. Circles, triangles and squares represent Fe-, Co- and Ir-based skutterudites. The black dashed line correspond to the $\frac{\kappa_P}{\kappa_C} = B^{-1}$ power law interpolating the data [see Eq. (B7)]. The shaded regions show a smooth crossover from a dominant particlelike heat conduction in green, to a competing particle- and wavelike mechanism in red. The burgundy cross represents the correlation for YbFe$_4$Sb$_{12}$ at 300 K. (Lower panel) Relation between the relative strength of particlelike and wavelike conduction ($\frac{\kappa_P}{\kappa_C}$) at 800 K and the descriptor $\eta$ given in Eq. (4). The linear correlation between $\eta$ and $B$ shows how skutterudites' chemistry determines the degree of Wigner thermal transport in these materials. The color scale goes from red (Wigner regime) to green (Boltzmann regime).

In the upper panel of Fig. 3 we show the correlation between $\frac{\kappa_P}{\kappa_C}$ and $B$ at 800 K. Interestingly, it can be shown analytically (see Appendix B) that, for a constant DOS, $\frac{\kappa_P}{\kappa_C} = B^{-1}$ and the upper panel of Fig. 3 shows computational proof of this. It is worth noting that the correlation for, e.g., YbFe$_4$Sb$_{12}$, at 300 K (burgundy cross in the upper panel of Fig. 3) shows how the relation $\frac{\kappa_P}{\kappa_C} = B^{-1}$ remains valid at different temperatures.

Finally, we want to understand how skutterudites' chemistry comes into play in discriminating the ability of the filler atom to move inside the cage and how this is related to $B$. The mean-square displacement (MSD) is often used in the literature to describe vibrating systems characterized by loosely bound atoms with long and elongated bonds [56] (see Appendix G for its formal definition). From this follows that we can define a heuristic parameter $\eta$ that captures how filler's vibrations fill the space available in the skutterudite's cage:

$$\eta = \left| \frac{\text{MSD}_R - \text{MSD}_{Sb}}{\text{MSD}_{Sb}} \frac{d_{Sb-R} - d_{Sb-Sb}}{d_{Sb-Sb}} \right|, \quad (4)$$

where $d_{Sb-R}$ is the distance between the filler atom and the atoms of the cage, $d_{Sb-Sb}$ is the bond length defining the Sb icosahedral cage where the filler is located and MSD$_R$ and MSD$_{Sb}$ are the mean-square displacement of the rattler and the Sb cage atoms, respectively. We note that, by definition, $\eta = 1$ for unfilled skutterudites.

In the lower panel of Fig. 3 we show the correlation between $\frac{\kappa_P}{\kappa_C}$ and $\eta$ at 800 K obtained for each of the filled skutterudites studied. This confirms the trend obtained for $\frac{\kappa_P}{\kappa_C}$ in the upper panel of Fig. 3, and validates numerically the predictions from Eqs. (2) and (B7). Most importantly, $\eta$ allows to connect the physics behind Wigner heat transport to the chemistry of the skutterudites: we find that $\eta$ correlates linearly with $B$. In this sense, $\eta$ can be used to provide a computationally cheap and close-to-being quantitatively accurate estimate of the degree of Wigner regime of thermal transport for a material. This underscores how $B$ is proportional to the rattling motion of the filler, quantified by $\eta$, and is therefore able to distinguish between optimal filler atoms for the reduction of $\kappa_{tot}$ from those for which the thermal behavior remains similar to that of unfilled skutterudites. For example, the filled skutterudite BaFe$_4$Sb$_{12}$ displays Boltzmann regime of thermal transport, since its $\eta$ is significantly lower than that of the other filled skutterudites. We also observe that a rescaling of the filler's atomic weight translates into negligible changes of $\kappa_{tot}$, thus confirming that thermal transport is determined by bonding chemistry (see Appendix E for details). In the end, it is worth noting that only the MSDs (harmonic properties) and the crystal chemical bonds enter the definition of $\eta$, and thus already at this level is it possible to give an approximate estimate of the degree of Wigner behavior of a thermoelectric material. Since the price of computing harmonic properties (MSDs) is orders of magnitude lower than that for computing anharmonic ones (phonon lifetimes), one could screen for cagelike thermoelectric materials with a strong wavelike contribution to conductivity through the parameter $\eta$. In practice, one may exploit this parameter to perform high-throughput computational screening of materials characterized by Wigner regime of thermal transport promising for thermoelectric applications.

## IV. CONCLUSION

In conclusion, we have used the Wigner formulation of thermal transport to investigate the microscopic physics underlying heat conduction in skutterudites, showing a crossover from Boltzmann to Wigner thermal transport when filling with, e.g., Yb atoms. Unfilled skutterudites display Boltzmann regime of thermal tranport, while filled ones change behavior from Boltzmann to Wigner regime depending on the filler atom and its bonding properties. We showed that, given the same host structure, the materials displaying the lowest conductivity are precisely those in which the $\frac{\kappa_P}{\kappa_C}$ ratio between particlelike and wavelike contributions is larger. We also elucidated how the degree of Wigner heat conduction (quantified by the Boltzmann deviation descriptor, $B$, derived from the microscopic harmonic and anharmonic quantities entering in the Wigner theory of thermal transport) is correlated to the relative motion between the filler atom and the cage; the latter being dependent on the chemical composition of the skutterudite structure (captured by the harmonic and computationally much cheaper parameter $\eta$). Thus, the rattling motion of the filler causing good thermoelectric performances can be seen as a direct manifestation of phonon coherences becoming as important as phonon population. Thereby, this study paves the way for the identification of the most suitable chemical compositions to engineer new and efficient cagelike thermoelectric materials.

## ACKNOWLEDGMENTS

This research was supported by the Swiss National Science Foundation (SNSF) through Grant No. CRSII5_189924 ("Hydronics" project). N.M. acknowledges support from NCCR MARVEL, a National Centre of Competence in Research, funded by the Swiss National Science Foundation (Grant No. 205602). M.S. acknowledges support from Gonville and Caius College, and from the SNSF project P500PT_203178.

## APPENDIX A: ANALYTIC DERIVATION OF A BOLTZMANN DEVIATION DESCRIPTOR

To derive a Boltzmann deviation descriptor ($B$) which captures the relative strength of the particlelike and wavelike heat conduction mechanisms, we start by resolving the phonon lifetime and total lattice thermal conductivity as a function of the frequency, $\tau(\omega)$ and $\kappa(\omega)$, respectively. In this way we are able to compare the filled skutterudites' properties with those of unfilled skutterudites, while keeping a unique theoretical scheme. To do this we exploit the Gaussian representation of the $\delta$ function to obtain the phonon lifetime and thermal conductivity as functions of frequency:

$$\tau(\omega) = \frac{\lim_{\epsilon \to 0^+} \sum_{q,s} \tau(q)_s \, e^{-\frac{(\omega(q)_s - \omega)^2}{2\epsilon^2}} \epsilon^{-1}}{\lim_{\epsilon \to 0^+} \sum_{q,s} e^{-\frac{(\omega(q)_s - \omega)^2}{2\epsilon^2}} \epsilon^{-1}}, \tag{A1}$$

$$\kappa(\omega) = \frac{\lim_{\epsilon \to 0^+} \sum_{q,s} \kappa(q)_s \, e^{-\frac{(\omega(q)_s - \omega)^2}{2\epsilon^2}} \epsilon^{-1}}{\lim_{\epsilon \to 0^+} \sum_{q,s} e^{-\frac{(\omega(q)_s - \omega)^2}{2\epsilon^2}} \epsilon^{-1}}. \tag{A2}$$

Within this frequency representation we want to determine how much each atom contributes to a certain phonon lifetime and energy. Such an atomic-resolved contribution can be obtained by using the partial density of states (PDOS) of the different types of atoms. In this way we can condense the information coming from the phonon frequency and phonon lifetimes spectra into two single scalars. It is worth noting that $\kappa(\omega)$ is used to keep track of the contribution that each phonon with frequency $\omega$ provides to the total conductivity. In such a way, the characteristic energy and lifetime expressions read as follows:

$$\hbar \bar{\omega}_i = \frac{\int_0^{\omega_{max}} \partial \omega \, \hbar \omega \, \kappa(\omega) \, \mathrm{PDOS}_i(\omega)}{\int_0^{\omega_{max}} \partial \omega \, \kappa(\omega) \, \mathrm{PDOS}_i(\omega)}, \tag{A3}$$

$$\tau_i = \frac{\int_0^{\omega_{max}} \partial \omega \, \tau(\omega) \, \kappa(\omega) \, \mathrm{PDOS}_i(\omega)}{\int_0^{\omega_{max}} \partial \omega \, \kappa(\omega) \, \mathrm{PDOS}_i(\omega)}, \tag{A4}$$

where $i = $ R, M, or Sb (M or Sb) represents the atom type for filled (unfilled) skutterudites. Since the Boltzmann or Wigner crystal behavior is regulated by the competition between phonon lifetime and Wigner limit in time [7] [see Eq. (2) of the main text], we can use the lifetimes resolved on the atoms

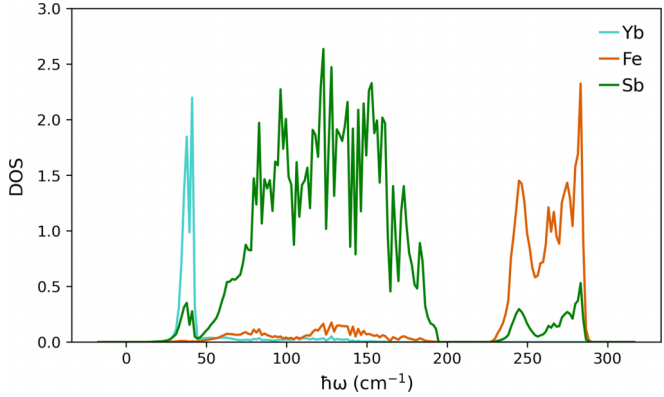

FIG. 4. PDOS of YbFe$_4$Sb$_{12}$.

and the average phonon interband spacing to define $B$:

$$B = \frac{1}{\bar{\tau}\,\Delta\omega_{\text{avg}}} = \begin{cases} \frac{1}{\tau^U\Delta\omega_{\text{avg}}}, & \text{with } \tau^U = \frac{\tau_M\tau_{Sb}}{\tau_M+\tau_{Sb}}, \\ \frac{1}{\tau^F\Delta\omega_{\text{avg}}}, & \text{with } \tau^F = \frac{\tau_R\tau_M\tau_{Sb}}{\tau_R\tau_M+\tau_R\tau_{Sb}+\tau_M\tau_{Sb}}, \end{cases}$$
(A5)

where the F and U superscripts symbolize filled and unfilled skutterudites, respectively. $B$ is devised to resolve, on average, how far the phonons are from the particle-wave crossover. It is worth noting, as shown in Fig. 2 of the main text, that it is not strictly necessary to obtain the average lifetime entering Eq. (A5) by discriminating the contributions resolved on the atoms (vertical coordinates of the black points in Fig. 2 of the main text) with the Matthiessen' sum. An identical result would also be obtained by mediating directly over all the phonon lifetimes (this is not true for the frequencies); in fact, the average lifetimes resolved on the atoms condense the information coming from the entire phonon lifetime cloud.

Finally, it is worth noting that the black symbols in Fig. 2 of the main text, whose $x$ coordinates are given by Eq. (A3), are displaying $y$ coordinates given by atom types resolved lifetimes [Eq. (A4)] which resemble the peaks in the PDOS of the related material (see Fig. 4). Therefore, we can conclude that our analysis based on the convolution of phonon modes $\boldsymbol{q}$, $s$ to their $\omega$-dependent analogous allows to correctly capture the lifetime of vibrations of the different atom types present in the structure.

## APPENDIX B: ANALYTIC RELATION BETWEEN $\frac{\kappa_P}{\kappa_C}$ AND A BOLTZMANN DEVIATION DESCRIPTOR

It is possible to give an analytical explanation for the interesting power relation, $\frac{\kappa_P}{\kappa_C} = B^{-1}$, obtained from Fig. 3 of the main text.

As discussed in the main text, the ratio between the wavelike and particlelike contributions to the lattice thermal conductivity is approximately equivalent to the ratio between the phonon linewidth and the average phonon interband spacing,

$$\frac{\kappa_C(\boldsymbol{q})_s}{\kappa_P(\boldsymbol{q})_s} \simeq \frac{\Gamma(\boldsymbol{q})_s}{\Delta\omega_{\text{avg}}} = \frac{1}{\tau(\boldsymbol{q})_s\Delta\omega_{\text{avg}}}.$$
(B1)

The above equation can be written as

$$\frac{\bar{\kappa}_P(\boldsymbol{q})_s}{\bar{\kappa}_C(\boldsymbol{q})_s}\frac{\delta(\omega_s(\boldsymbol{q})-\bar{\omega})}{\delta(\omega_s(\boldsymbol{q})-\bar{\omega})} \simeq \tau(\boldsymbol{q})_s\Delta\omega_{\text{avg}}\frac{\delta(\omega_s(\boldsymbol{q})-\bar{\omega})}{\delta(\omega_s(\boldsymbol{q})-\bar{\omega})},$$
(B2)

where $\delta(\omega_s(\boldsymbol{q})-\bar{\omega})$ is the Dirac $\delta$ and $\bar{\omega}$ is the average frequency of the crystal obtained by averaging the $\bar{\omega}_i$ given in Eq. (A3), where $i$ = R, M, or Sb (M or Sb) represents the atom type for filled (unfilled) skutterudites. Since Eq. (B2) holds $\forall\boldsymbol{q}$ we can integrate numerator and denominator of both sides over the Brillouin zone:

$$\frac{\frac{1}{(2\pi)^3}\int_{BZ}\partial\boldsymbol{q}\,\bar{\kappa}_P(\boldsymbol{q})_s\,\delta(\omega_s(\boldsymbol{q})-\bar{\omega})}{\frac{1}{(2\pi)^3}\int_{BZ}\partial\boldsymbol{q}\,\bar{\kappa}_C(\boldsymbol{q})_s\,\delta(\omega_s(\boldsymbol{q})-\bar{\omega})}$$

$$\simeq \frac{\frac{1}{(2\pi)^3}\int_{BZ}\partial\boldsymbol{q}\,\tau(\boldsymbol{q})_s\Delta\omega_{\text{avg}}\delta(\omega_s(\boldsymbol{q})-\bar{\omega})}{\frac{1}{(2\pi)^3}\int_{BZ}\partial\boldsymbol{q}\,\delta(\omega_s(\boldsymbol{q})-\bar{\omega})}.$$
(B3)

Given the definition of $\Delta\omega_{\text{avg}}$, where it is assumed that the phonon bands are uniformly spaced, from the previous equation we obtain the expression

$$\frac{\bar{\kappa}_P(\omega)}{\bar{\kappa}_C(\omega)} \simeq \frac{\tau(\omega)\Delta\omega_{\text{avg}}}{\text{DOS}(\omega)}$$
(B4)

that holds $\forall\omega$. We can then write

$$\frac{\int_0^{\omega_{max}}\partial\omega\,\bar{\kappa}_P(\omega)}{\int_0^{\omega_{max}}\partial\omega\,\bar{\kappa}_C(\omega)} \simeq \frac{\Delta\omega_{\text{avg}}\int_0^{\omega_{max}}\partial\omega\,\tau(\omega)}{\int_0^{\omega_{max}}\partial\omega\,\text{DOS}(\omega)}$$

$$= \frac{\Delta\omega_{\text{avg}}\int_0^{\omega_{max}}\partial\omega\,\tau(\omega)\text{DOS}(\omega)}{\int_0^{\omega_{max}}\partial\omega\,\text{DOS}(\omega)}$$

$$\times \frac{\int_0^{\omega_{max}}\partial\omega\,\text{DOS}(\omega)}{\int_0^{\omega_{max}}\partial\omega\,\text{DOS}^2(\omega)}.$$
(B5)

In this way we have

$$\frac{\bar{\kappa}_P}{\bar{\kappa}_C} \simeq \bar{\tau}\,\Delta\omega_{\text{avg}}\frac{\int_0^{\omega_{max}}\partial\omega\,\text{DOS}(\omega)}{\int_0^{\omega_{max}}\partial\omega\,\text{DOS}^2(\omega)}.$$
(B6)

If we assume DOS($\omega$) to be a constant we simply end with

$$\frac{\bar{\kappa}_P}{\bar{\kappa}_C} \simeq \bar{\tau}\,\Delta\omega_{\text{avg}} \to \frac{\bar{\kappa}_P}{\bar{\kappa}_C} \simeq B^{-1}.$$
(B7)

## APPENDIX C: COMPUTATIONAL DETAILS

### 1. Approximations to the exchange-correlation energy functional

The equilibrium crystal configuration of the skutterudites studied were obtained using the Quantum ESPRESSO (QE) distribution [57] through Kohn-Sham density functional theory (DFT) [58,59] calculations with projector-augmented-wave (PAW) [60] (for Ba, Nd, Yb, I, In, La, Fe) and ultrasoft (US) (for Ca, Co, Ir) pseudopotentials, as suggested by the standard solid-state pseudopotentials (SSSP) library [61–63]. We employed the generalized gradient approximation (GGA), formulated by the Perdew-Burke-Ernzerhof (PBE) [64] exchange-correlation functional for all compounds. This conclusion was reached after testing the LDA, PBE and PBEsol [65] functionals with respect to the agreement with the cell parameter measured experimentally.

TABLE I. Optimized structure parameters ($a_{calc.}$) of unfilled $FeSb_3$, $CoSb_3$, and $IrSb_3$ and of their related filled compounds $RFe_4Sb_{12}$ (R = Ba, Ca, Nd, Yb), $RCo_4Sb_{12}$ (R = I, In, La, Yb), and $RIr_4Sb_{12}$ (R = Yb) compared with the available experimental data ($a_{expt.}$) given alongside. 1, 2, and 3 superscripts refer to an experimental concentration of the filler equal to 0.8, 0.23, and 0.3 per unit cell, respectively.

| Material | $FeSb_3$ | $BaFe_4Sb_{12}$ | $CaFe_4Sb_{12}$ | $NdFe_4Sb_{12}$ | $YbFe_4Sb_{12}$ | $CoSb_3$ | $ICo_4Sb_{12}$ | $InCo_4Sb_{12}$ | $LaCo_4Sb_{12}$ | $YbCo_4Sb_{12}$ | $IrSb_3$ | $YbIr_4Sb_{12}$ |
|---|---|---|---|---|---|---|---|---|---|---|---|---|
| $a_{calc.}$ | 9.209 | 9.236 | 9.191 | 9.179 | 9.178 | 9.119 | 9.203 | 9.198 | 9.226 | 9.129 | 9.392 | 9.407 |
| $a_{expt.}$ | 9.212 [67] | 9.201 [34] | 9.159 [34] | 9.136 [34] | 9.159 [34] | 9.055 [68] | 9.122 [37][1] | | 9.060 [40][2] | 9.063 [43][3] | 9.250 [44] | |

Before the geometry optimization, the convergence of total energy, forces and pressure with respect to the $k$-mesh, kinetic energy cutoff and value of the spreading (smearing) for brillouin-zone integration was achieved using a $6 \times 6 \times 6$ grid, 60 Ry and 0.02 Ry, respectively. We employed the Marzari-Vanderbilt smearing [66]. The convergence thresholds used to obtain the parameters entering the DFT calculations are given by 0.05 Ry variation of the total energy, 1 kbar variation of total pressure and 0.001 Ry/Bohr variation of total force.

### 2. Structures optimization

The geometry optimization of the unit cell of all the compounds studied was carried out by using the set of converged parameters discussed in the previous section. The calculated lattice parameters ($a_{calc.}$) of the materials studied together with the related experimental values ($a_{expt.}$) are summarized in Table I. Overall, the obtained values are in good agreement with the available experimental data, apart from the Ir-based skutterudites where the unit cell size is slightly overestimated.

### 3. Interatomic force constants calculations

The harmonic interatomic force constants (IFCs) needed to evaluate the scattering term in the Wigner transport equation were calculated through the `Quantum ESPRESSO PHonon` package by means of density functional perturbation theory (DFPT) [69] calculations on a $3 \times 3 \times 3$ Monkhorst-Pack grid.

We generated supercell structures with random atomic displacements by exploiting the `hiPhive` [70] code. This tool relies on compressive sensing methods that efficiently construct sparse solutions for linear systems (which in the present case reflect the short-range nature of IFCs). The ordinary least-squares optimization was used to provide solutions of the linear problem. The third-order IFCs were calculated through DFT evaluation of atomic forces acting on displaced atoms by using a $2 \times 2 \times 2$ supercell with a $3 \times 3 \times 3$ $k$ grid.

### 4. Thermal properties calculations

The phonon scattering rates and the lattice thermal conductivity were calculated with the `Phono3py` [71] package which solves the Wigner transport equation with the SMA approximation. We took into account the scattering effects due to anharmonicity (up to the three-phonon interaction) and to isotopic disorder.

Furthermore, we use the `SHENGBTE` [72] code (and its `thirdorder.py` auxiliary tool) to benchmark the results obtained with the `hiPhive+Phono3py` workflow discussed above.

The lattice thermal conductivity calculations were performed using the smearing method through the `Quantum ESPRESSO D3Q` package [6,73–75] and the tetrahedron method [76,77] implemented in Phono3py [71].

In the following paragraph we discuss the method used to obtain the optimized parameters needed for deriving the thermal properties of the materials studied. Again, we show the convergence study for $YbFe_4Sb_{12}$, taken as a prototype structure. We analyze the differences in terms of lattice thermal conductivity results due to the different methods used to derive the IFCs discussed so far.

### 5. Convergence with respect to calculation parameters

A systematic study of the convergence for lattice thermal conductivity calculations of $YbFe_4Sb_{12}$ with respect to the number of `hiPhive` input atomic configurations is summarized in Fig. 5. All the calculations in Fig. 5 rely on the second-order force constants obtained through `Quantum ESPRESSO` DFPT calculations, as they are more precise and less expensive than the supercell ones; as known, this is due to the translational invariance that allows to work in reciprocal space within the Brillouin zone only. Moreover, all the calculations were carried out using `D3Q` package with the smearing method. We then focus on the calculations performed with `SHENGBTE (thirdorder.py)`. For these it is necessary to specify only the cutoff value of the third order of interaction since, as we have already said, the second-order force constants were obtained with DFPT. The cutoff definition in `SHENGBTE` is given by the maximum distance between two atoms over which they are considered as noninteracting. So, this means that before starting the calculation, one has to choose up to which level the interaction is considered (which reflects the precision one wants to achieve) and then, by varying it, see when the result of the lattice thermal conductivity calculation does not depend on it. The reference value for our calculations is given by considering the third order of interaction in `SHENGBTE` up to the sixth nearest neighbor atom. In this case, the result obtained is the one given by the red curve in Fig. 5; this is because, given the interaction range defined by the number of nearest neighbor atoms to be considered, `SHENGBTE` automatically evaluates the related cutoff value in angstroms.

Also, once specified the number of nearest neighbors, `SHENGBTE`, returns the atomic positions configurations needed to implement the small displacement method. The number of these input configurations is therefore strictly related to the value of the cutoff (or number of nearest neighbors) that one chooses. In this sense, therefore, we have a cutoff of 7.9 Å (red curve) when considering the interaction up to six nearest neighbors, 7.7 Å (dark green curve) for five nearest neighbors and 6.0 Å (brown curve) for three nearest neighbors. As it can be seen again from Fig. 5, the red and dark green curves, in addition to having the same trend, are also very close to each other, meaning that considering the interaction up to five nearest neighbors is already enough to achieve convergence for `SHENGBTE` calculations. From what has just been discussed, it is now clear why it was decided to take the red curve calculation as the reference one.

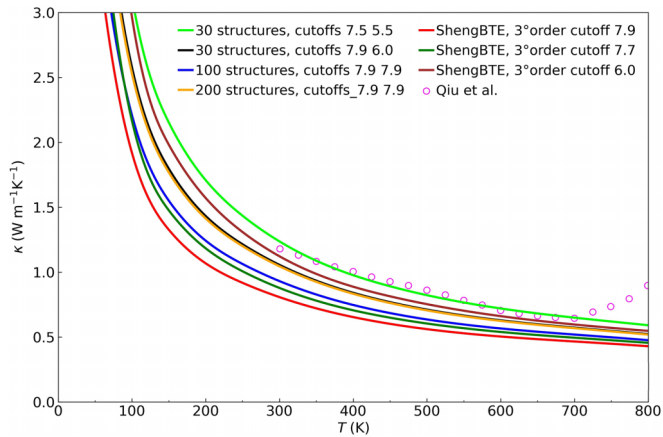

FIG. 5. Convergence of the lattice thermal conductivity of YbFe$_4$Sb$_{12}$ with respect to the number of input atomic configurations and the third- and second-order cutoffs of interaction used for the `hiPhive` calculation and with respect to the third-order cutoff of interaction used for the `SHENGBTE` calculation. The first column of the legend refers to the calculations performed with `hiPhive`, in fact both the number of input structures (atomic positions configurations randomly generated) and the cutoffs of interactions are specified. The second column instead contains the calculations carried out with `SHENGBTE` and the experimental reference data. Note that all the calculations were carried out using the second-order force constants obtained through `Quantum ESPRESSO` DFPT calculations. For `hiPhive` calculations the first value of the cutoffs is the one of the second order of interaction while the second is the one of the third order of interaction. Instead, for `SHENGBTE` calculations, the only cutoff value that appears is clearly the one of the third order of interaction. The cutoff values are measured in angstroms (Å). The experimental values are taken from Ref. [34].

Now we analyze in detail the convergence of `hiPhive` calculations. As can be seen in the first column of the legend of Fig. 5, for `hiPhive` calculations it was necessary to specify also the cutoff value of the second order of interaction (given that the second-order force constants obtained from DFPT are still the used ones). In fact, `hiPhive` needs the values of the cutoffs on both orders of interaction to be able to correctly fit both the second and third-order force constants, which are obtained simultaneously. This means that the convergence should no longer be studied on the single value of the cutoff of the third order of interaction, but on the configuration of cutoff values of second and third order of interaction together.

Also, in `hiPhive` calculations, the number of input structures is specified *a priori*. Therefore, as can be seen from Fig. 5, the used number of structures is also a parameter for which the convergence has to be studied. In fact, the used number of `hiPhive` structures, and therefore the number of self-consistent calculations needed, is far less than those required in a `SHENGBTE` calculation. For example, to perform a `SHENGBTE` calculation with the interaction up to the sixth nearest neighbor, it was necessary to perform 1292 self-consistent calculations. As will be clearer below, a result very similar to the one obtained in this way is deduced through 100 self-consistent calculations where the input structures were generated with `hiPhive`. The difference between these two methods is that in the first case the number of self-consistent calculations depends on the specified cutoff value of the interaction while in the second case this is not true; this is why convergence must also be tested with respect to the number of input structures.

Furthermore, since `hiPhive` implements a compressive sensing method, also the definition of cutoff is slightly different from that provided for `SHENGBTE`: the cutoff in `hiPhive` is defined as the maximum distance between two atoms (over which they are considered as noninteracting) within the same cluster, where cluster simply stands for a subset of lattice points. The size of a cluster (commonly referred to as the cluster radius) is defined as the average distance to the geometrical center of the cluster. `hiPhive` works by decomposing the supercell into clusters of points in the lattice. Then it defines a set of symmetry equivalent clusters that allows to reduce the computational cost of the calculation as many elements of the force constant matrices are related to each other by symmetry operations. Finally, the irreducible set up of parameters is obtained by applying all symmetry operations allowed by the space group of the ideal lattice and the acoustic sum rules.

As it is easy to see from Fig. 5, the second-order cutoff is always greater than or equal to the third-order cutoff. This is intuitively due to the fact that pair interactions have longer range than many-body interactions, and so second- and third-order cutoffs have to be chosen consistently as to avoid overfitting problems.

The idea behind the convergence test in the `hiPhive` calculation lies in choosing the second-order cutoff as extreme as possible, i.e., the maximum value that the code can handle before encountering double clusters and so aborting the fitting procedure. In the study of Fig. 5, this maximum value for the second-order cutoff is 7.9 Å. Note that this value has nothing to do with the number of input structures used in the calculation but depends exclusively on the size of the system, i.e., the size of the supercell.

Once the value of the second-order cutoff has been fixed, it is possible to study how the lattice thermal conductivity results change as the value of the third-order cutoff (chosen to be equal or smaller than that of the second order) varies. To avoid overfitting, it is possible to consider a greater number of input structures or to reduce the demand of cutoff values. In fact, the black and blue lines in Fig. 5, representing, respectively, the calculation with 30 structures, second-order cutoff value of 7.9 Å and third-order cutoff value of 6.0 Å, and the calculation with 100 structures, second-order cutoff value of 7.9 Å and third-order cutoff value of 7.9 Å, both return a result close to the reference red curve.

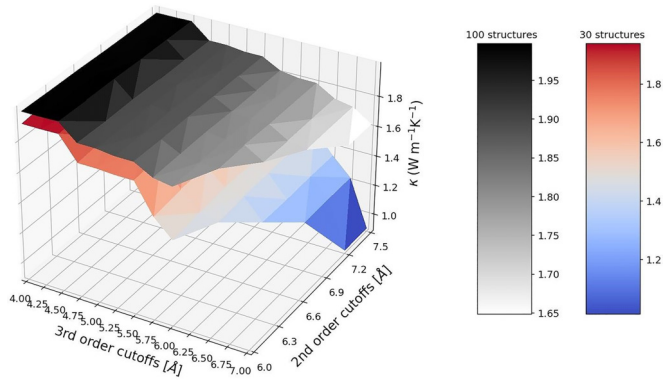

FIG. 6. Convergence of the lattice thermal conductivity of a generic skutterudite with respect to the number of input configurations and the cutoff values of the second and third order of interaction.

It can therefore be said that the calculations relating to the black and blue curves in Fig. 5 correctly describe the same physics. It is worth mentioning that the variation of the third-order cutoff value between these two calculations was not selected randomly. In fact, for all the materials studied, it was possible to find a trend in lattice thermal conductivity as a function of cutoff values and number of structures like the one summarized in Fig. 6. Although from Fig. 6 it can be clearly seen that the lattice thermal conductivity keeps varying as the cutoff values change, what is important to underline is that the surface formed by the calculation on the 100 structures is much flatter than the one obtained from the 30 structures. This suggests that, once a configuration of cutoff values is fixed, as the number of the input structures increases, an increasingly flat surface is obtained in which therefore the conductivity varies less and less. This behavior is explained as follows: increasing the initial data set ensures greater accuracy of the fitting method, but clearly the big gain in computational savings would be lost.

However, Fig. 6 provides additional information. Given the same second-order cutoff value, the conductivity result calculated on the 100 structures is very similar to that calculated on the 30 structures, in which the third-order cutoff value is lower than that used for the 100 structures. To clarify what has just been said it may be useful to focus on two particular pairs of points in the plot of Fig. 6. For example, for the pairs of points (with fixed second-order cutoff value equal to 6.0 Å) given by (6.0 Å, 4.50 Å) for the 30 structures calculation and (6.0 Å, 4.75 Å) for the 100 structures calculation, we can see roughly the same value of the lattice thermal conductivity. Note that we have defined these points where the first coordinate is the second-order cutoff value and the second coordinate is the third-order cutoff value. Again this can be shown also for the pairs of points given by (6.0 Å, 5.50 Å) for the 30 structures calculation and (6.0 Å, 5.75 Å) for the 100 structures calculation, and so on. In analogy to this deduction, the optimal parameters configuration for a small number of input structures (30 structures) in Fig. 5 was given by second- and third-order cutoff values equal to 7.9 Å and 6.0 Å, respectively.

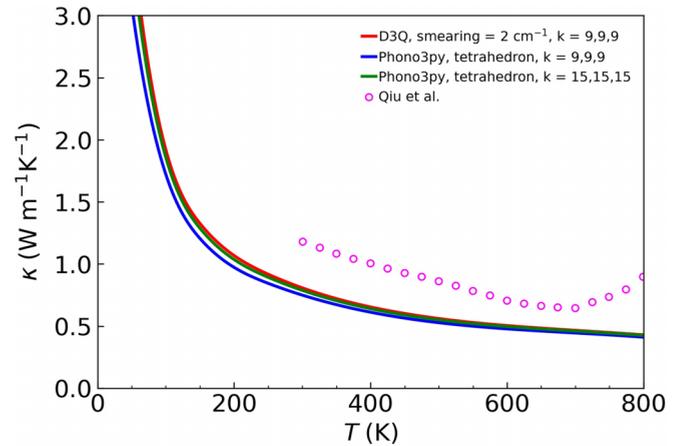

FIG. 7. Comparison of `D3Q` with smearing method and `Phono3py` with tetrahedron method calculations in YbFe$_4$Sb$_{12}$. The used smearing value used is measured in $Wm^{-1}K^{-1}$ and the $k$-points grid has also been specified. The experimental values are taken from Ref. [34].

## 6. Comparison between different calculation methods

Figure 7 summarizes the lattice thermal conductivity calculations carried out using `D3Q` package with smearing method and `Phono3py` software with tetrahedron method. It can be clearly seen that, regardless of the $k$-mesh used, the two different `Phono3py` calculations almost perfectly match the reference one, obtained by using `D3Q` package with smearing method (red curve in both Figs. 5 and 7). In particular, this technology ends up to be less expensive and faster because it makes a strong use of symmetries and calculates the lattice thermal conductivity in many fewer points than `D3Q` does.

Note that smearing and $k$-points mesh values were subjected to convergence: by varying the smearing from 2 to 4 or 6 cm$^{-1}$ and the mesh from $9 \times 9 \times 9$ to $11 \times 11 \times 11$, $13 \times 13 \times 13$, or $15 \times 15 \times 15$ the final results did not change.

Thanks to the convergence tests and the comparison of different calculations of the lattice thermal conductivity, it was possible to provide a configuration of parameters and methods to be used for the calculation of the lattice thermal conductivity for all the materials studied; this configuration is summarized in Table II.

TABLE II. Computational methods and parameters used for the lattice thermal conductivity calculations of the materials studied.

| Software package | Number of structures | Cutoffs values (Å) | Integration method | $k$-mesh |
|---|---|---|---|---|
| `hiPhive` | 30 | 7.9 and 6.0 | Tetrahedron method | $9 \times 9 \times 9$ |

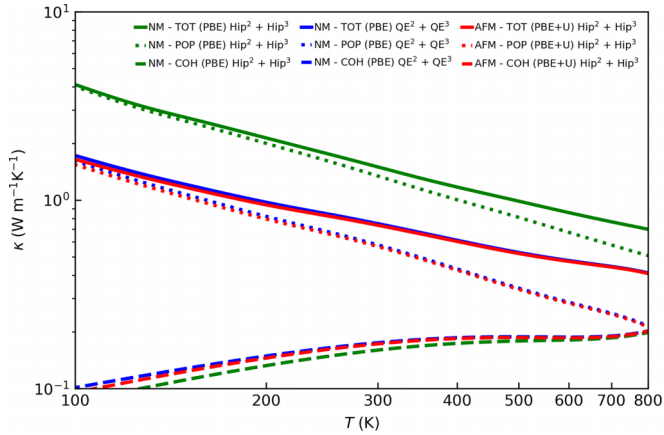

FIG. 8. Total lattice thermal conductivity (solid), particle-like contribution (dotted), and wavelike contribution (dashed) of YbFe₄Sb₁₂ obtained from DFT calculations on top of `hiPhive` displaced atomic position configurations (green) for both second- and third-order interatomic force constants, `PHonon` DFPT calculations for second-order force constants, and DFT calculations on top of `hiPhive` displaced atomic position configurations for third-order force constants (blue) and DFT + U calculations on top of `hiPhive` displaced atomic position configurations (red) for both second- and third-order interatomic force constants. Here "2" and "3" superscripts refer to the order of the interatomic force constants. A Hubbard parameter U = 2.5 eV was used for the DFT + U calculations. Note that the PBE + U + hiPhive calculation derives from the related antiferromagnetic electronic state with lowest energy, while the PBE+ Quantum ESPRESSO DFPT and PBE + hiPhive ones derive from the related nonmagnetic electronic state with lowest energy.

### 7. Phonon scattering processes considered in the Wigner transport equation

To solve the WTE [Eq. (1) of the main text] we take into account anharmonic and isotopes scattering processes through the standard collision operator as defined in Ref. [73]. We do not consider the boundary scattering as we are referring to large samples whose finite size effects of the crystal are negligible [73]. In the kinetic regime of thermal transport [74], where Umklapp scattering events dominate, and so where the lattice thermal conductivity ($\kappa_{tot}$) becomes very low at medium-high temperatures (as in the case of filled skutterudites), a good estimate of the particlelike conductivity is given by the single-mode approximation (SMA) [11,73]. Solving the WTE with this approximation allows to write $\kappa_{tot}$ in the following compact form given in Eq. (1) of the main text.

### APPENDIX D: MAGNETIC ORDER AND DFT+HUBBARD

The nonmagnetic phase of FeSb₃ displays imaginary phonon frequencies, as already highlighted in other works [32,78]. Therefore, we follow Ref. [32] where an antiferromagnetic (AFM) ground state was obtained for FeSb₃ in the primitive unit cell and the experimental paramagnetic phase of this material was addressed through the special quasirandom structure (SQS) method [79], which is used to maximize magnetic moments disorder [80,81] minimizing the total energy in supercell representation. In the aforementioned work a DFT+Hubbard [29–31,82–85] approach was used to correct the self-interaction error (SIE) [33,86] of partially occupied iron $d$ orbitals due to approximate forms of DFT.

We chose to benchmark the problem by studying YbFe₄Sb₁₂ with the DFT + Hubbard functional (DFT + U) and allowing for magnetic orders. Also for YbFe₄Sb₁₂ an AFM ground state has been obtained. It should be noted that the calculation of the second-order interatomic force constants is practically not feasible using DFPT. This is so because at the first iteration of the resolution of the linear system set up for the computation of the force constants it is necessary to converge the bare (noninteracting) response function $\chi_0$ which requires a huge computational power.

What we show here is that the thermal properties, such as the lattice thermal conductivity, do not vary appreciably if instead a nonmagnetic ground state without Hubbard corrections is considered. This analysis therefore confirmed that magnetism in skutterudites plays a fundamental role but does not particularly affect thermal transport. In fact, by looking at Fig. 8 we can clearly see how the reference DFT result that we have discussed so far, that is, the calculation obtained through the use of the DFPT second-order force constants and the `hiPhive` third-order ones (blue), is completely analogous to that obtained through DFT + U and allowing for AFM order (red).

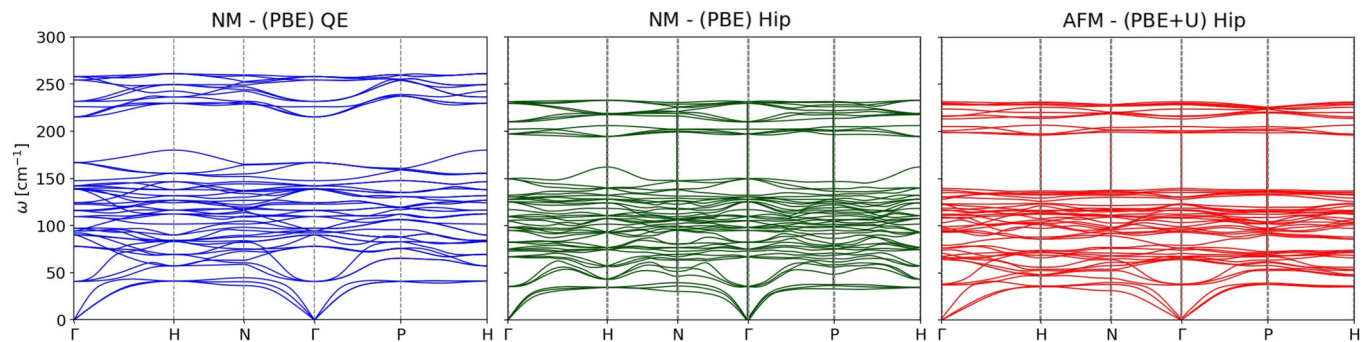

FIG. 9. Phonon dispersions of nonmagnetic YbFe₄Sb₁₂ obtained using PBE+Quantum ESPRESSO DFPT (blue), antiferromagnetic YbFe₄Sb₁₂ with PBE+U+hiPhive (red) and nonmagnetic YbFe₄Sb₁₂ with PBE+hiPhive (green).

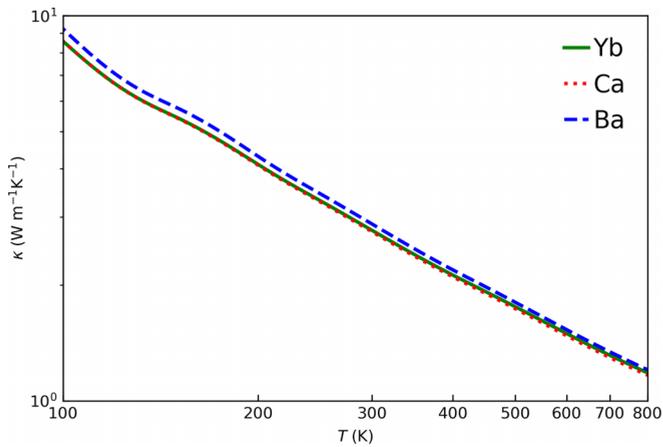

FIG. 10. Total lattice thermal conductivity of $BaFe_4Sb_{12}$ obtained by substituting the atomic weight of Ba with that of Yb (solid green), by substituting the atomic weight of Ba with that of Ca (dotted red) and by using the actual atomic weight of Ba (dashed blue).

The phonon dispersions obtained from the second-order force constants used for the lattice thermal conductivities of Fig. 8 are shown in Fig. 9. We clearly see how the highest computational level we can reach with the DFT+Hubbard theory returns phonons similar to those considered in the text.

## APPENDIX E: EFFECT OF FILLER ATOMIC WEIGHT SUBSTITUTION

We want to highlight once more that the complexity of the physics underlying heat transport in skutterudites is due to the chemistry of the various elements that compose them. The trend of the thermal conductivity of $BaFe_4Sb_{12}$ (the filled Fe-based skutterudite showing the smallest wave-tunneling contribution to thermal transport) starting from the interatomic force constants calculated by replacing the atomic weight of Ba (137.327 amu), respectively, with that of Yb (with atomic weight close to Ba, 173.045 amu) and that of Ca (with much lower atomic weight than Ba, 40.078 amu) is given in Fig. 10. It is evidently seen that the difference in atomic weight does not appreciably affect the lattice thermal conductivity. We clearly observe how the substitution of the filler's atomic weight does not influence the thermal conductivity, showing that a rescaling of the filler's atomic weight translates into negligible changes of $\kappa_{tot}$, and thus confirming that thermal transport is determined by the bonding chemistry.

## APPENDIX F: VIBRATIONAL PROPERTIES OF THE SKUTTERUDITES STUDIED

### 1. Phonon dispersions of the skutterudites studied

In Fig. 11 we summarize the phonon dispersions of the skutterudites studied. As can be easily seen, for all the materials we do not observe imaginary phonon frequencies.

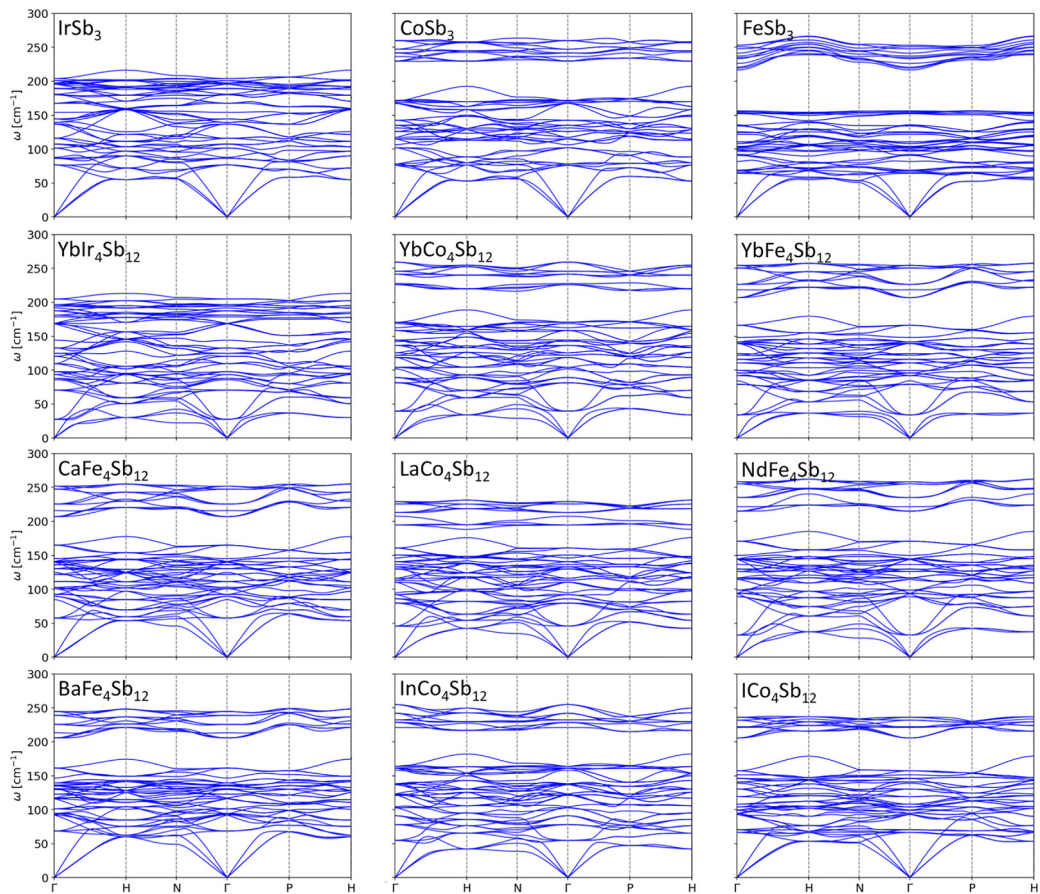

FIG. 11. Phonon dispersions of the skutterudites studied. Note that the phonon dispersion of $FeSb_3$ comes from DFPT calculations carried out on top of the antiferromagnetic electronic ground state, while for the other materials the nonmagnetic ground state was considered.

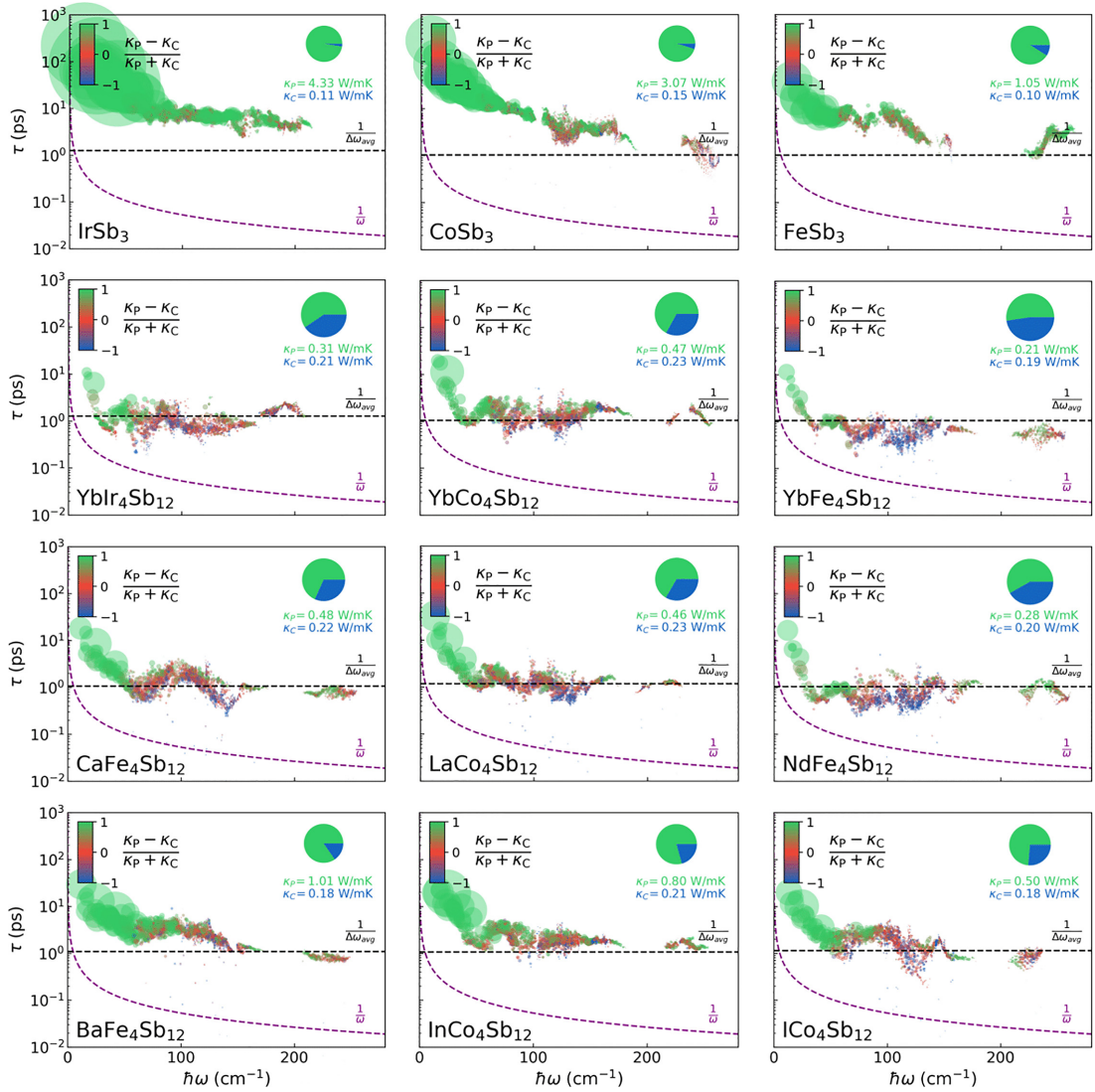

FIG. 12. Phonon lifetimes distribution $\tau(\boldsymbol{q})_s = \frac{1}{\Gamma(\boldsymbol{q})_s}$ as a function of the energy $\hbar\omega(\boldsymbol{q})_s$ for the skutterudites studied at 800 K.

### 2. Phonon lifetimes distribution of the skutterudites studied

In Fig. 12 we summarize the phonon lifetimes distributions of the skutterudites studied.

### 3. Phonon lifetimes distribution of the YbFe$_4$Sb$_{12}$ at 300 K, 800 K, and 1200 K

To quantify what happens at different temperatures, we look at the phonon lifetimes vs energy plots at $T = 300$, 800, and 1200 K (Fig. 13). We see that for $T \gtrsim 800$ K the Wigner-crystal behavior is increasingly important, with coherences conductivity becoming larger than population's conductivity.

### APPENDIX G: MEAN-SQUARE DISPLACEMENT DEFINITION

The mean-square displacement (MSD) entering Eq. (4) of the main text can be defined also for crystals following Eq. (10.71) of Ref. [87]. The atomic displacement, $\boldsymbol{u}$, is

written as

$$u^\alpha(jl,t) = \sqrt{\frac{\hbar}{2Nm_j}} \sum_{\boldsymbol{q},s} [\omega_s(\boldsymbol{q})^{-\frac{1}{2}} (\hat{a}_s(\boldsymbol{q})e^{-i\omega_s(\boldsymbol{q})t} + \hat{a}_s^\dagger(-\boldsymbol{q})e^{i\omega_s(\boldsymbol{q})t})e^{i\boldsymbol{q}\cdot\boldsymbol{r}(jl)}\gamma_s^\alpha(j,\boldsymbol{q})], \quad \text{(G1)}$$

where $j$ and $l$ are the labels for the $j$th atomic position in the $l$th unit cell, $t$ is the time, $\alpha$ is an axis, $m$ is the atomic mass, $N$ is the number of the unit cells, $\boldsymbol{q}$ is the wave vector, $s$ is the index of phonon mode. $\gamma$ is the polarization vector of the atom $jl$ and the band $s$ at $\boldsymbol{q}$. $\boldsymbol{r}(jl)$ is the atomic position and $\omega$ is the phonon frequency. Finally, $\hat{a}$ and $\hat{a}^\dagger$ are the annihilation and creation operators of phonon. The expectation value of the squared atomic displacement is then calculated as [88]

$$\langle |u^\alpha(jl,t)|^2 \rangle = \frac{\hbar}{2Nm_j} \sum_{\boldsymbol{q},s} \omega_s(\boldsymbol{q})^{-1} (1+2n_s(\boldsymbol{q},T)) |\gamma_s^\alpha(j,\boldsymbol{q})|^2, \quad \text{(G2)}$$

where $n_s(\boldsymbol{q},T)$ is the phonon population given by the Bose-Einstein distribution.

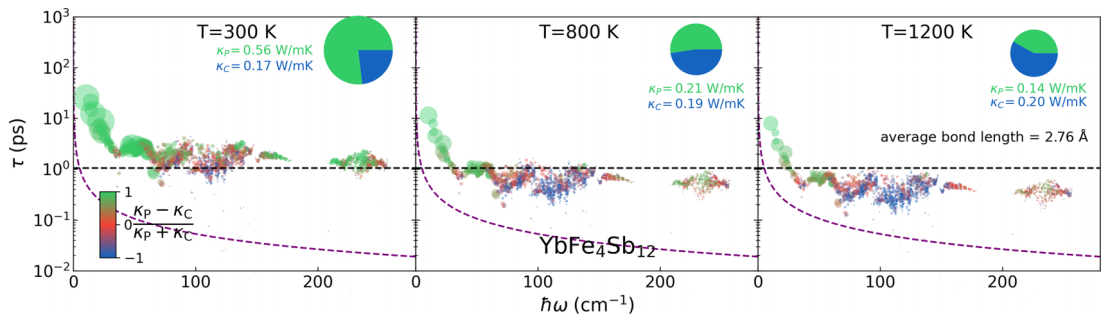

FIG. 13. Distribution of phonon lifetimes $\tau(\boldsymbol{q})_s = \Gamma(\boldsymbol{q})_s^{-1}$ as a function of energy $\hbar\omega(\boldsymbol{q})_s$ for $YbFe_4Sb_{12}$ at 300 K (left), 800 K (center), and 1200 K (right). We see that the wavelike contribution reaches a plateau at 800 K as we do not have further appreciable increase of $\kappa_C$ up to 1200 K.